\documentclass[]{aa}
\usepackage[varg]{txfonts}
   \usepackage[flushleft]{threeparttable}
\usepackage{graphicx}
\usepackage{subfig}
\usepackage{natbib}

\title{Impacts of fragmented accretion streams onto Classical T
Tauri Stars: UV and X-ray emission lines}

\author{S. Colombo\inst{\ref{inaf}} \and S. Orlando\inst{\ref{inaf}}
\and G. Peres\inst{\ref{inaf},\ref{unipa}} \and
C. Argiroffi\inst{\ref{inaf},\ref{unipa}} \and
F. Reale.\inst{\ref{inaf}, \ref{unipa}}}

\institute{INAF-Osservatorio Astronomico di Palermo, Piazza del
Parlamento 1, 90134 Palermo, Italy \label{inaf} \and Dipartimento
di Fisica \& Chimica, Università degli Studi di Palermo, Piazza del
Parlamento 1, 90143 Palermo, Italy \label{unipa}}
\newcommand\salvo[1]{{#1}}

\begin{document}

\abstract
{\salvo{The accretion process in Classical T Tauri Stars (CTTSs)
can be studied through the analysis of some UV and X-ray emission
lines which trace hot gas flows and act as diagnostics of the
post-shock downfalling plasma. In the UV band, where higher spectral resolution is available, these lines are
characterized by rather complex profiles whose origin is still not
clear.}}
{\salvo{We investigate the origin of UV and X-ray emission at impact
regions of density structured (fragmented) accretion streams. We
study if and how the stream fragmentation and the resulting structure
of the post-shock region determine the observed profiles of UV and
X-ray emission lines.}}
{We model the impact of \salvo{an accretion stream consisting of a
series of dense blobs} onto the chromosphere \salvo{of a CTTS
through} 2D \salvo{MHD} simulations. We explore different
\salvo{levels of stream fragmentation and accretion rates. From the
model results, we synthesize C\,IV (1550 \AA) and O\,VIII (18.97 \AA)
line profiles.}}
{\salvo{The impacts of accreting blobs onto the stellar
chromosphere produce reverse shocks propagating through the blobs and shocked upflows. These upflows, in turn, hit and shock the subsequent downfalling fragments. As a result, several plasma components
differing for the downfalling velocity, density, and temperature are present altoghether. The profiles of C\,IV doublet are characterized by two main components: one narrow and redshifted to speed $\approx 50$~km s$^{-1}$ and the other broader and consisting of subcomponents with redshift to speed in the range $200 - 400$~km s$^{-1}$. The profiles of O\,VIII lines appear more symmetric than C\,IV and
are redshifted to speed
$\approx 150$~km s$^{-1}$.}}
{\salvo{Our model predicts profiles of C\,IV line remarkably similar to those observed and explains their origin in a natural way as due to stream fragmentation.} }

\maketitle

\section{Introduction}

Classical T Tauri Stars (CTTS) are young stars surrounded by \salvo{an
accretion} disk. \salvo{According to the magnetospheric accretion
scenario, gas from the disk accretes onto the star through accretion
columns \citep{1991Apj...370L..39K}. Several lines of evidence
support this picture especially in the optical and infrared bands
(e.g. \citealt{1988ApJ...330..350B}).} Also accreting T Tauri Stars show a
\salvo{soft X-ray} (0.2-0.8 KeV) excess, with typical lines produced
at \salvo{temperatures} $10^5-10^6$ K. \salvo{This excess has been
interpreted as due to impacts of the accreting material with the
stellar surface where a shock is produced and dissipates the kinetic
energy of the downfalling material \citep{1991Apj...370L..39K}. The
shock heats the plasma up to temperatures of few million degrees,
causing X-ray emission \citep{2002ApJ...567..434K, 2007A&A...465L...5A}.
The heated plasma is characterized by density of $n\approx 10^{11}-10^{13}$
cm$^{-3}$ (e.g. \citealt{2007A&A...465L...5A}).}

\salvo{The interpretation of the soft X-ray excess in CTTSs in terms
of accretion shocks is well supported by hydrodynamic (HD) and
magnetohydrodynamic (MHD) models.} 
{Time-dependent 1D model of radiative accretion shocks in CTTSs provided a first accurate description of the dynamics of the postshock plasma \citep{2008MNRAS.388..357K,2008A&A...491L..17S}. In particular \cite{2008A&A...491L..17S} proposed a one.dimensional hydrodynamic (HD) model of a continuous accretion flow impacting onto the chromosphere of a CTTS, thus assuming the ratio between the thermal pressure and the magnetic pressure $\beta <<1$. Their model reproduced the main features of high spectral resolution X-ray observations of the CTTS MP Mus that was previously interpreted as due to postshock plasma \citep{2007A&A...465L...5A} }

\salvo{More recently 2D MHD models of accretion impacts have been
studied (\citealt{2010A&A...510A..71O, 2013A&A...557A..69M,
2013A&A...559A.127O}) to explore those cases where the low-$\beta$
approximation cannot be applied (and, therefore, the 1D models
cannot be used). These models have shown that the accretion dynamics
can be complex with the structure and stability of the impact region
of the stream strongly depending on the configuration and strength
of the stellar magnetic field. Depending on the magnetic field
strength, the atmosphere around the impact region can be also
perturbed, leading to accreting plasma leaks at the border of the
main stream.}

\salvo{Although HD and MHD models of accretion shocks have provided
a theoretical support to the hypothesis that the soft X-ray excess
in CTTSs originates from impacts of accretion columns onto the
stellar surface, several points still remain unclear. Some of these
points concern the emission in the UV band arising from impact
regions. There is} evidence that \salvo{a significant amount of
plasma at $10^{5}$ K (much larger than expected from current models;
Costa et al. 2016, submitted)} \salvo{is} produced \salvo{in the accretion process}.
\cite{2013ApJS..207....1A} analyzed UV spectra collected with the
{\it Hubble Space Telescope} (HST) of 28 T Tauri stars and studied
the C\,IV doublet at 1550 \AA. They found that each component of
the doublet is described by 2 Gaussian components with different
speed and width. \salvo{About half of their sample exhibits line
profiles analogous (but with different Doppler shifts) to those of TW Hya that consist of a narrow
component redshifted at speeds of $\approx 30$ km s$^{-1}$ (positive
speed indicates material that falls into the star) and a broader
component centered at $\approx 120$ km s$^{-1}$ and with the
redshifted wing reaching $\approx 400$ km s$^{-1}$. The latter
component cannot be explained, as post-shock emission, with current models of a
continuous accretion stream: assuming a free fall velocity of
$\approx 500-600$ km s$^{-1}$ and a strong shock, the velocity of
the post-shock plasma cannot be larger than $\approx 100-120$ km
s$^{-1}$ \citep{2010A&A...522A..55S}.}

\salvo{Recently \cite{2014ApJ...797L...5R} proposed an explanation
on the origin of the observed asymmetries and redshifts of UV
emission lines in CTTSs. These authors studied the impacts of dense
plasma fragments falling back on the surface of the Sun after a
violent eruption occurred on 2011 June 7, showing that this phenomenon
reproduces on the small scale accretion impacts onto CTTSs (see
also \citealt{2013Sci...341..251R}). They modeled the impacts with
HD simulations and synthesized the emission in UV and X-ray bands.
They found that UV emission may originate from the shocked front
shell of the still downfalling fragments, thus producing a broad
redshifted component in UV lines up to speeds around $\approx 400$
km s$^{-1}$.}

\salvo{In this work we investigate further the scenario of a
fragmented stream through MHD modeling. More specifically we study
the structure and stability of the post-shock plasma after the
impact of a clumpy or fragmented stream onto the stellar surface.
We investigate the origin of UV and X-ray emission at impact regions
and if and how the stream fragmentation can be responsible of the
observed asymmetries and redshifts of UV emission lines in CTTSs.
To this end, we developed an MHD model describing an accretion
column consisting of several high density blobs which impact onto
the chromosphere of a CTTS.  We synthesized the C\,IV (1550 \AA)
and O\,VIII (18.97 \AA) emission lines, including the effect of
Doppler shift due to plasma motion along the line of sight.
The paper is structured as follow: in Sect.~\ref{sec2} we describe the
model and the synthesis of emission lines; in Sect.~\ref{sec3} we
discuss the results of the simulations and the synthesis of emission
lines; and finally in Sect.~\ref{sec4} we drawn our conclusions.}

\section{The Model}
\label{sec2}

The model describes a fragmented accretion stream impacting onto
the surface of a CTTS. We assume that the accretion occurs along
magnetic field lines that link the circumstellar disk to the surface
of the star, and that the accretion stream is not continuous but
is composed by blobs with different density.

Our model takes into account the stellar magnetic field, the gravity,
the radiative cooling from optically thin plasma and the thermal
conduction, including the effects of heat flux saturation. The
impact of the accretion stream is modelled by solving the time-dependent
MHD equations:

\begin{gather}
	\frac{\partial}{\partial t}\rho + \nabla \cdot \rho \vec{v} = 0\\
	\frac{\partial}{\partial t}\vec{m}+\nabla \cdot (\vec{m}\vec{v} - \vec{B}\vec{B}+ \vec{I}p_t) = \rho\vec{g}\\
	\frac{\partial}{\partial t}(E)+\nabla\cdot( (E+p_t)\vec{v}-\vec{B}(\vec{v}\cdot\vec{B}))=\vec{m}\cdot\vec{g} + \nabla \cdot F_c - n^2 {\Lambda}(T)	\label{RL}\\
	\frac{\partial}{\partial t}\vec{B}+\nabla\cdot(\vec{v}\vec{B}-\vec{B}\vec{v})=0
\end{gather}

\noindent
where, $\rho$ is the density, $\vec{v}$ the plasma velocity,
$\vec{m}=\rho \vec{v}$ the momentum in volume unit, $\vec{B}$ the
magnetic field, $p_t=p+B^2/2$ the total pressure (magnetic and
thermal), $E$ is the total energy density ($E=\rho \epsilon +
\frac{1}{2}\rho v^2 + \frac{1}{2}B^2$), $\vec{g}$ is \salvo{the}
gravity, $F_c$ is \salvo{the} conductive flux, $n$ is the plasma density and ${\Lambda}(T)$
represents the optically thin radiative losses per unit emission
measure derived with the PINTofALE \citep{2000HEAD....5.2705K}
spectral code with the CHIANTI atomic lines database 
\citep{2013ApJ...763...86L} using solar abundances.


{In the presence of the organized stellar magnetic field,
the thermal conduction is anisotropic (in particular it is known
to be highly reduced in the direction transverse to the field) and
can be locally split into two components, along and across the
magnetic field lines: $F_c=F_{||}i+F_{\bot}j$. To allow for a smooth
transition between the classical and saturated conduction regime,
we followed \cite{1993ApJ...404..625D}. The thermal fluxes along and
across the magnetic field lines are}

\begin{gather*}
F_{||} = \left(\frac{1}{[q_{spi}]_{||}}+\frac{1}{[q_{sat}]_{||}}\right)^{-1}\\
F_{\bot} = \left(\frac{1}{[q_{spi}]_{\bot}}+\frac{1}{[q_{sat}]_{\bot}}\right)^{-1}
\label{thermal_conduction}
\end{gather*}
	
\noindent
{where $[q_{spi}]_{||}$ and $[q_{spi}]_{\bot}$ represent the
classical conductive flux along and across the magnetic field lines
according to \cite{1962pfig.book.....S}:}

\begin{gather*}
[q_{spi}]_{||} =-k_{||}[\nabla T]_{||} = -9.2 \cdot10^{-7}T^{5/2}[\nabla T]_{||}\\
[q_{spi}]_{\bot}= -k_{\bot}[\nabla T]_{\bot} = -5.4\cdot10^{-16}n^2/(T^{1/2}B^2)[\nabla T]_{\bot}
 \end{gather*}

{$k_{||}$ and $k_{\bot}$ are both in units of $\text{erg}
\cdot \text{K}^{-1} \text{s}^{-1}\text{cm}^{-1}$ and $[\nabla
T]_{||}$ and $[\nabla T]_{\bot}$ are the thermal gradients along
and across the magnetic field lines. For temperature gradient scales
comparable to the electron mean free path, the heat flux is limited
and the conductive flux along and across the magnetic field lines
are given by \citep{1977ApJ...211..135C}:}

\begin{gather*}
[q_{sat}]_{||} = -\text{sign}([\nabla T]_{||}) = 5\phi\rho c^3_{\text{S}}\\
[q_{sat}]_{\bot} = -\text{sign}([\nabla T]_{\bot}) = 5\phi\rho c^3_{\text{S}}
\label{cond_sat}
\end{gather*}

\noindent
{where $ c_{\text{S}} $ is the isothermal sound speed, $\rho$ is the
plasma density and $\phi$, called flux limit factor, is a free
parameter between 0 and 1 \citep{1984ApJ...277..605G}. For this
work $\phi= 0.9$.}

\begin{figure}
	\centering
	\includegraphics[scale=0.5]{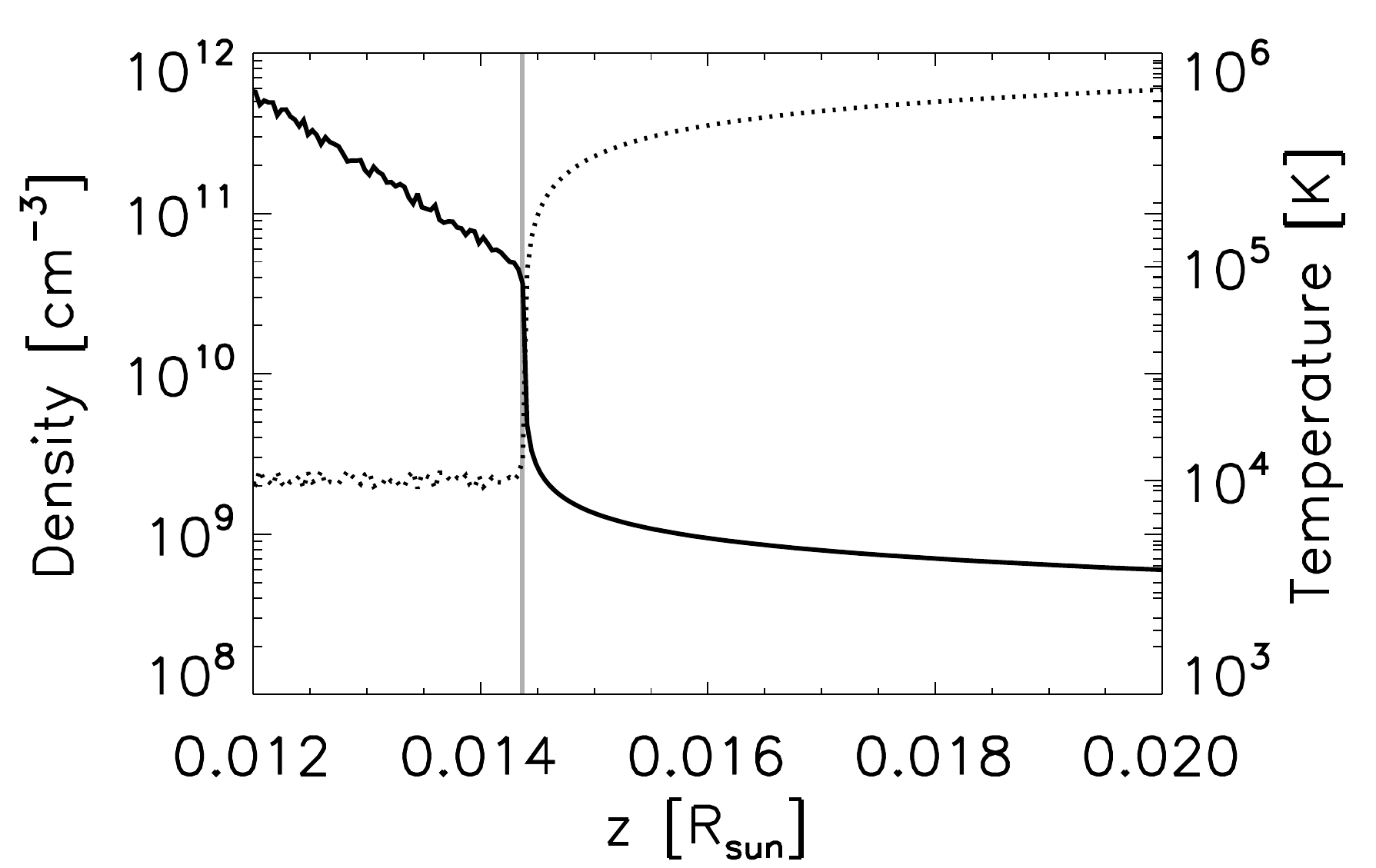}
	\caption{Enlargement of initial \salvo{vertical profiles of} density (continuous line)
	and temperature (dotted line) in logarithmic scale. The
	vertical grey line \salvo{shows the transition region between
	the chromosphere and the corona}.\label{CI}}
\end{figure}

We adopted \salvo{a} cylindrical geometry and \salvo{solved} the
MHD \salvo{equations} in the plane \salvo{$(r,z)$}. The left side
of the domain is the axis of symmetry. Initially the domain describes
a stellar isothermal chromosphere that ends at $\approx 0.014 R_{\odot}$
\salvo{linked through a steep transition region to a corona expanding up
to the end of the domain \citep{1996JGR...10124443O}. The atmosphere is hydrostatic and plane-parallel. We introduced per-cell random density perturbations in the
whole domain, with maximum perturbations of 10\% of theoretical
density values.} The \salvo{initial} magnetic field is 500 G and is
uniform in the whole domain and perpendicular to the surface of the
star. The gravity is calculated assuming $M_{\text{star}}=1.2M_\odot$
and $R_{\text{star}}=1.3R_\odot$ and it is uniform in the whole
domain. Fig. \ref{CI} shows the initial conditions.

\salvo{The accretion stream is assumed to fall along the $z$
axis with a velocity of about $500  $ km s$^{-1}$ when it impacts onto the
chromosphere. The stream is in pressure equilibrium with the surrounding medium and consists
of a column with density $10^9$~cm$^{-3}$ and of a train of consecutive
blobs with higher density. The density of the blobs is a free
parameter that we change in the simulations. We also performed a simulation describing a continuous accretion
stream to provide a baseline case for the simulations of fragmented
streams.}

Table \ref{table} summerizes the parameters of the simulations:
model abbreviation, blob density ($\rho_{bl}$), interblob density ($\rho_{str}$), impact speed ($V_{imp}$), time between two consecutive blobs ($\Delta t_{bl}$), blob length ($R_{bl}$) and accretion rate ($\dot{M}_{acc}$).

\salvo{In addition to the simulations describing the impact of a
train of blobs, we performed simulations describing the more general
case of a falling series of circular fragments (blobs) with a random
spatial distribution. We assumed the stream consisting of a column
with density $10^{9}$~cm$^{-3}$ and a series of blobs with random
values of density ranging between $5\times10^{10}$~cm$^{-3}$ and
$5\times10^{11}$~cm$^{-3}$.  For the runs Frag-N20 and Frag-N55 the Table \ref{table} reports the number of blobs described in the initial conditions ($N_{bl}$). For these simulations we adopted a
cartesian coordinate system and solved the MHD equations in the
plane $(x,y)$. These simulations are analogous to those presented
by \cite{2014ApJ...797L...5R} except for the presence of the stellar
magnetic field which we considered here. This allowed us to compare
our results with those presented in \cite{2014ApJ...797L...5R} and
to evaluate the role of the magnetic field in determining the
structure of the post-shock plasma.}

\begin{table*}[!htbp]
	\caption{Parameters of the simulations \label{table}}
	\centering
	\begin{tabular}{ccccccc}
		\hline
		\hline
		Simulation\tablefootmark{a} & $\rho_{bl}$          & $\rho_{str}$       & $V_{imp}$ & $\Delta t_{bl}$ &$R_{bl}$& $\dot{M}_{acc}$ \\ 
		             & ($10^{11}$cm$^{-3}$) &  ($10^9$cm$^{-3}$) & $10^7$cm s$^{-1}$&     (s)         &R$_{\odot}$& $10^{-11} M_{\odot} yr^{-1}$ \\	
		\hline		
	TR-D11-T300-R10.3           &   1                  & 1                 & 5 &  300           & 0.27& 10.3  \\
	  TR-D11-T150-R10.3        &   1                  & 1                 & 5 &  150           & 0.27&  9.3  \\  
  TR-D11-T600-R10.3         &   1                  & 1                 & 5 &  600           & 0.27&  6.14 \\ 
  TR-D11.7-T300-R10.3           &   5                  & 1                 & 5 &  300           & 0.27& 50.3  \\  	
  TR-D11-T300-R9.9         &   1                  & 1                 & 5 & 300           &  0.135&  5.9  \\  
  1BL-R10.3           &   1                  & 1                 & 5 & ---          & 0.27&  8.6  \\ 
	    CONT         &   ---              & 100               & 5 &  0             & --- & 22.5    \\  
	    \hline
	  Simulation \tablefootmark{a} & $\rho_{bl}$          & $\rho_{str}$	     & $V_{imp}$ & $ N_{bl}$       & $R_{bl}$ &$\dot{M}_{acc}$ \\
			                 &                      &                    &           &					&  R$_{\odot}$  &$10^{-11} M_{\odot} yr^{-1}$\\
	  \hline
	   Frag-N20       & 0.5-5                & 1                  & 5 & 20              &    0.005 - 0.02       & $0.6$   \\
 	   Frag-N55       & 0.5-5                & 1                  & 5 & 55              &    0.005 - 0.01      & $0.55$  \\
		\hline
		\end{tabular}

	\tablefoot{\tablefoottext{a}{Impact speed is $\approx 500 $ km s$^{-1}$.}}

\end{table*}

The calculations were performed using PLUTO v4.1
\citep{2007ApJS..170..228M}, a modular, Godunov-type code for
astrophysical plasmas. The code \salvo{is designed} to use parallel computers
using Message Passage Interface (MPI) libraries.  The MHD equations
are solved using the MHD module available in PLUTO with the
Harten-Lax-van Leer Discontinuities (HLLD) approximate Riemann
solver \citep{2005AGUFMSM51B1295M}. The evolution of the magnetic
field is calculated using the constrained transport method \salvo{\citep{1999JCoPh.149..270B}} that maintains the
solenoidal condition at machine accuracy. The time evolution is
solved using a second order Runge-Kutta method. The radiative losses
${\Lambda}$ are calculated at the temperature of interest using a
table lookup/interpolation method. The thermal conduction is treated
with a super time-stepping method; the superstep consists of N
substeps, properly chosen for optimization and stability, depending
on the diffusion coefficient, the grid size and the free parameter
$\nu < 1$ \citep{CNM:CNM950}.

The \salvo{2D} domain consists of an uniform grid with a cell size
of \salvo{$2.15\times10^8$ cm} on the \salvo{horizontal coordinate
($r$ in cylindrical coordinates and $x$ in cartesian) and
$2.69\times10^6$ cm on the vertical coordinate ($z$ in cylindrical,
$y$ in cartesian)}. At the lower boundary we set
conditions for density, pressure and temperature to describe the
stellar chromosphere\salvo{; a discontinuous inflow is defined at
the upper boundary. In simulations adopting a cylindrical geometry
we assumed axisymmetry on the left boundary ($r=0$) and outflow on
the right boundary; in simulations adopting a cartesian geometry
we used periodic boundary conditions in both the left and right
boundaries.}

\subsection{Synthesis of UV and X-ray emission}
\label{sec:synth}

\salvo{From the model results, we synthesized the C\,IV (1550 \AA)
and O\,VIII (18.97 \AA) emission lines, both belong to a doublet. First we reconstructed the
3D spatial distributions of plasma density, temperature, and velocity
by rotating the 2D spatial domain around the axis of symmetry (i.e.
the z axis). We calculated the emission measure of the j-th cell
as $EM_{j}(T)=\rho_{j}dV_{j}$, where $dV_{j}$ is the volume of the
cell. Then the distribution of emission measure vs. temperature,
EM$(T)$, is derived by binning the emission measure values as a
function of temperature in the range $[4.5 < \log T ({\rm K}) <
7]$; the range of temperature is divided into 30 bins, all equal
on a logarithmic scale. From the distributions of emission measure
and temperature, we synthesized UV and X-ray spectra, using
the CHIANTI atomic lines database \citep{2013ApJ...763...86L} and assuming solar metal
abundances. The spectral synthesis includes the Doppler shift of
lines due to plasma motion along the line of sight, we supposed that the angle between the line of sight and the axis of accretion column is 0. We integrated
the X-ray and UV spectra from the cells in the whole spatial domain. For
each simulation, we summed the spectra derived for the different
time frames and divided the resulting spectra by the total time of
the simulation. Finally we convolved our high resolution spectra
with a Gaussian function of $\sigma$ 17  km s$^{-1}$, for C\,IV profiles, to approximate the spectral
resolution of HST observations (\citealt{2013ApJS..207....1A}). For O\,VIII profiles we convolved our high resolution spectra with a Gaussian function of $\sigma$ of 81 km s$^{-1}$ to approximate the spectral resolution of Chandra/HETG \footnote{http://cxc.harvard.edu/proposer/POG/html/chap8.html}.}

\section{Results}
\label{sec3}

\subsection{\salvo{Impact of a train of blobs}}
\label{sec:train}

\salvo{We considered as a reference the case of a stream consisting
of a train of blobs aligned one to the other with density $\rho =10^{11}$cm$^{-3}$, downfalling
velocity of $\approx 500$ km s$^{-1}$, and with a time interval
between two consecutive blobs of $\Delta t = 300$~s(run TR-D11-T300-R10.3 in Table \ref{table}). The density
of the interblob medium is $10^9$~cm$^{-3}$ in all the simulations.
In order to study the effects of stream fragmentation, we
performed also a simulation describing a continuous stream with
density $\rho = 10^{11}$~cm$^{-3}$ for comparison.}

\begin{figure*}
\centering
\includegraphics[scale=0.75]{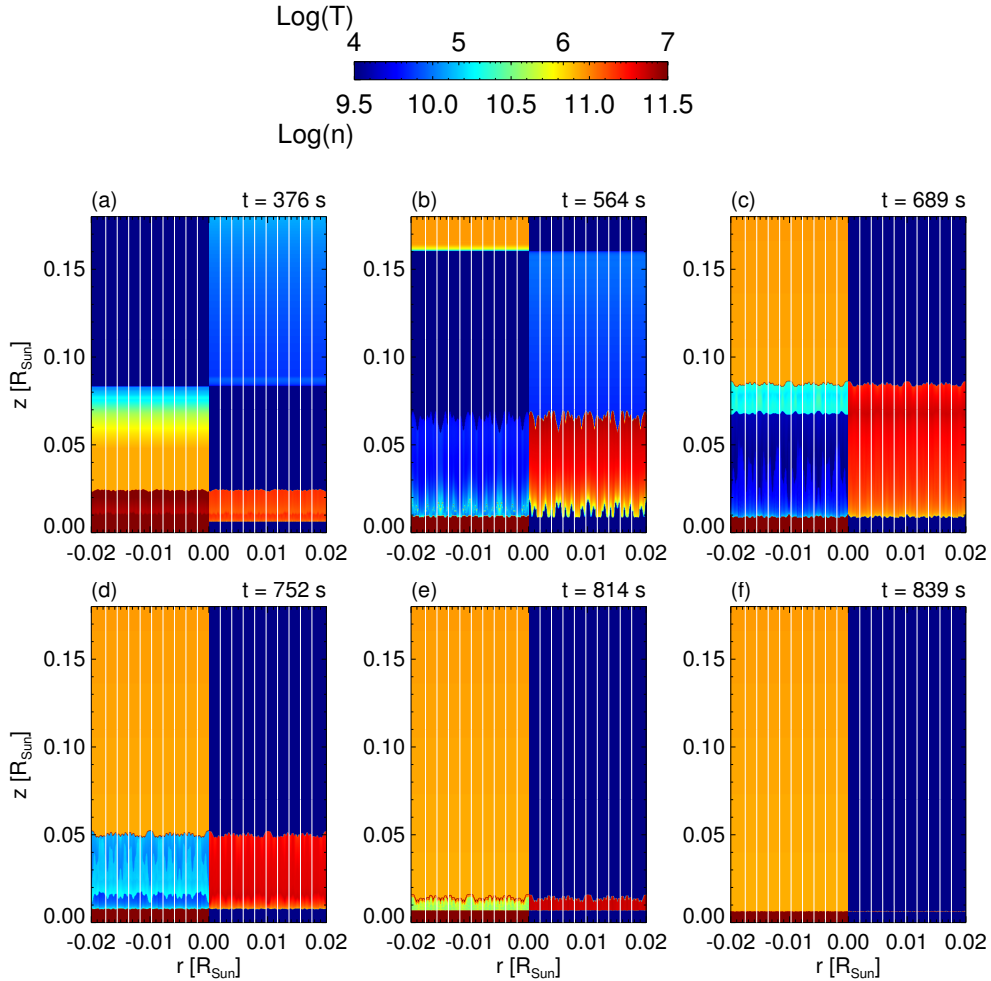}

       \caption{Color maps in log scale of evolution of density
 (left \salvo{half-panel}) and temperature (right \salvo{half-panel}) of plasma for the
 reference case (run TR-D11-T300-R10.3 in Table\ref{table}). White lines represent magnetic field lines.\label{sim_ref}}

\end{figure*}

\salvo{Initially the evolution of the train of blobs is analogous
to that of the continuous stream. Fig.~\ref{sim_ref} shows maps of
temperature and density for our reference case in a time range
between 5 and 15 minutes from the beginning of the simulation.
Movies showing the complete evolution of 2D spatial distributions
of mass density (on the left) and temperature (on the right) in log
scale are provided as online material. The train of blobs flows
along the magnetic field lines, and the first blob impacts the
stellar surface at $t \approx 50$~s at about 500  km s$^{-1}$
(Fig.~\ref{sim_ref}a) and sinks into the chromosphere until its ram
pressure equals the thermal pressure of the chromosphere.}
		
The collision generates a transmitted \salvo{and} a reverse shock,
the \salvo{latter travels back through the blob and} gradually
\salvo{builds up a dense slab of shock-heated plasma at temperatures
of few MK (Fig \ref{sim_ref}b). Given the magnetic field strength
($500$~G) considered in our simulations, the plasma $\beta << 1$
in the slab. As a consequence, mass and energy exchanges across
field lines are prevented and the slab is structured in several
fibrils, each independent of the others \citep{2010A&A...522A..55S}.
The evolution of each of these fibrils consists of alternating
phases of expansion and collapse of the shock-heated plasma (see
\citealt{2010A&A...522A..55S}). The expansion phase is guided by
the reverse shock and ends when the post-shock plasma becomes
thermally unstable. Then the system enters in the collapse phase:
the catastrophic cooling robs the post-shock plasma of pressure
support, causing the collapse of the material above. After the
fibril has disappeared in chromosphere, a new fibril is re-built
by the reverse shock. {During the evolution, the thermal
conduction continuously drains energy from the shock-heated plasma
to the chromosphere, thus acting as an additional cooling mechanism
of the hot slab (see also \cite{2010A&A...510A..71O}). At the same time,
the conduction contrasts the radiative cooling. The heat flux
saturation (see Eqs. \ref{cond_sat}) limits the effects of thermal
conduction where temperature gradients are very steep, in particular
at the contact discontinuity between the chromosphere and the hot
slab and in regions affected by catastrophic cooling of dense plasma.}

The evolution of the train of blobs departs from that of the
continuous stream when the reflected shock reaches the upper boundary
of the blob (Fig.~\ref{sim_ref}b). At that time, the pressure of
the shocked blob is much higher than the ram pressure of the layers
above due to the low density of the interblob region. As a result
the hot and dense plasma of the shocked blob expands adiabatically
upward through the accretion column. During this phase the shock-heated
plasma gradually cools down as a result of the adiabatic expansion.
The expansion ends when the upflowing surge} collides with the
following blob generating a second shock region high in the accretion
column (h=0.1$R_{\odot}$) (Fig.~\ref{sim_ref}d, e). In this phase
the blob plasma \salvo{can reach temperatures up to $10^7$ K}. This
shock region is dragged down by the \salvo{blob at velocities of
$\approx 500$ km s$^{-1}$ until it impacts the chromosphere and
sinks} (Fig. \ref{sim_ref}f). \salvo{Then a new shock develops and
travels through the blob, re-building the hot slab.}

\begin{figure}
	\centering
		\includegraphics[]{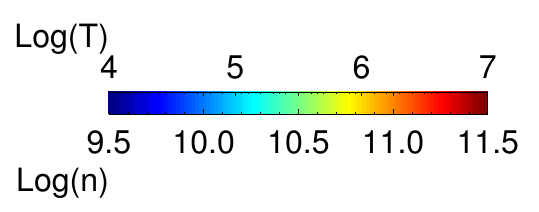}
	\centering
	
		\includegraphics[scale=0.58]{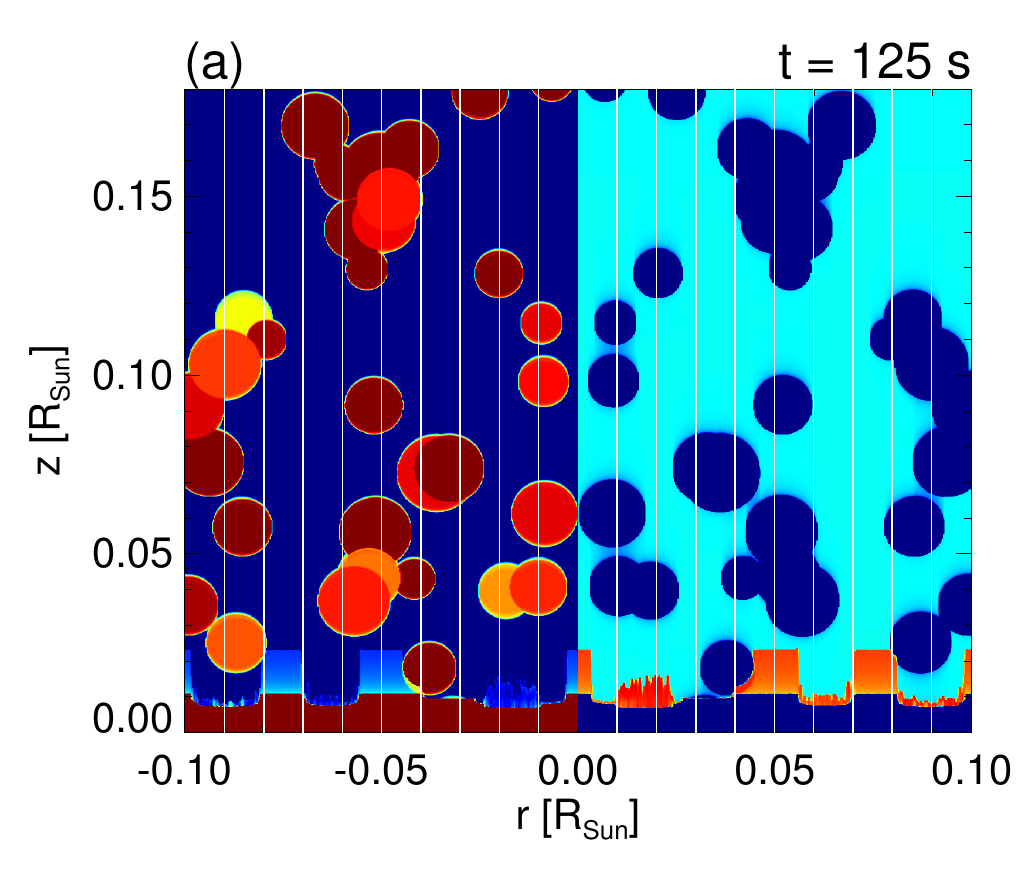}\\

		\includegraphics[scale=0.58]{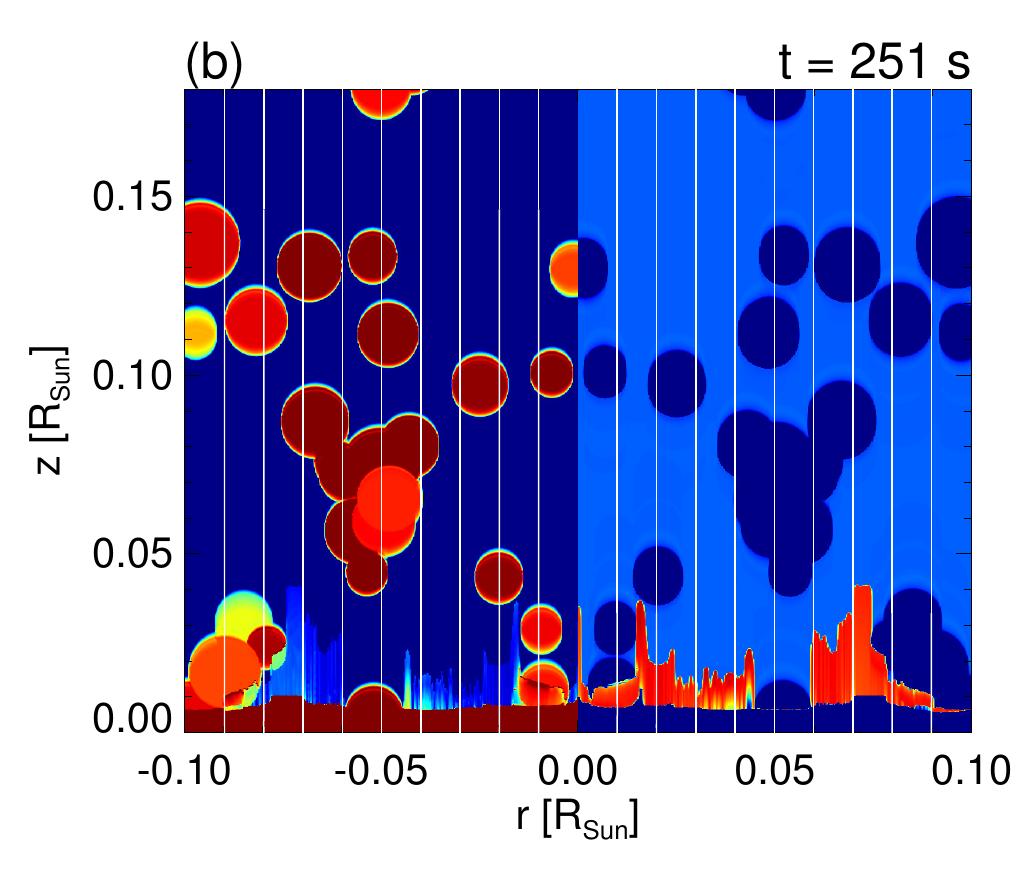}\\

		\includegraphics[scale=0.58]{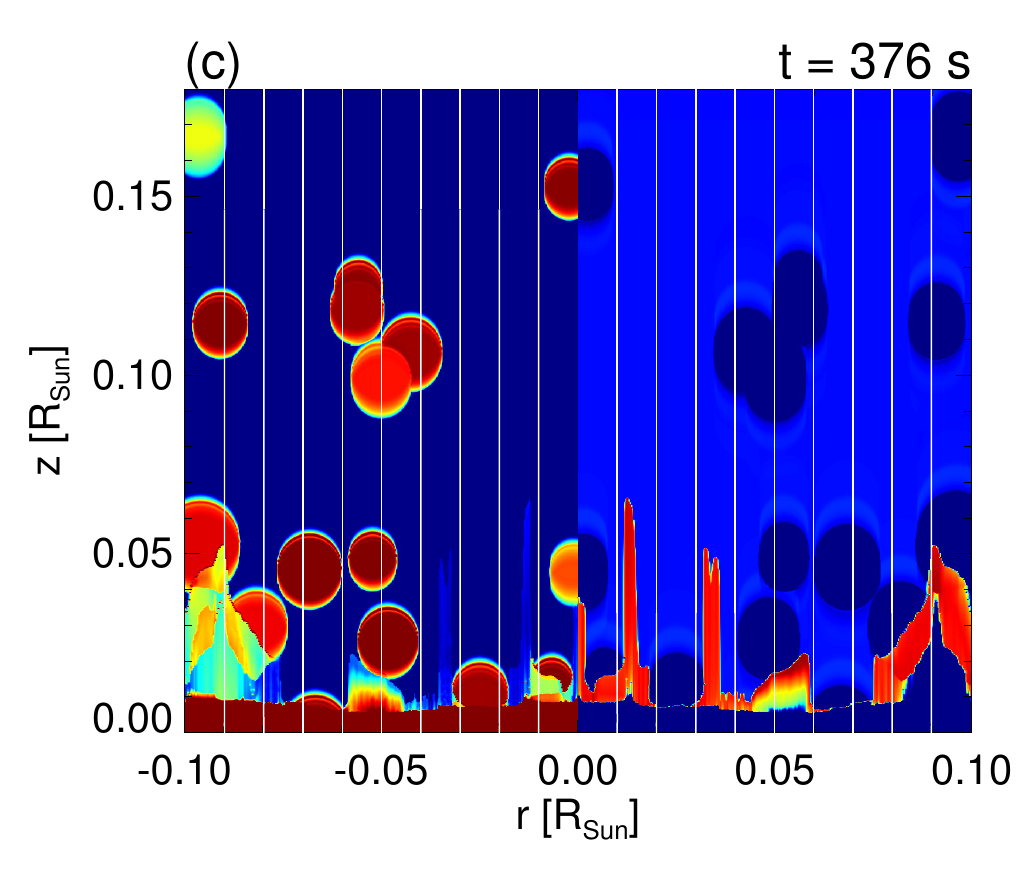}\\
	
	\caption{Color maps in log scale of evolution of density
		(left \salvo{half-panel}) and temperature (right \salvo{half-panel}) of plasma for the general case of fragmented stream (run Frag-55 in Table \ref{table}). White lines represent magnetic field lines.\label{sim_frag}}
	
\end{figure}

\subsection{Exploration of the parameter space}

\salvo{For the model describing a train of blobs, we performed an exploration of the parameter space defined by three free parameters (see Table~\ref{table}): the frequency
of blobs (or, in other words, the distance between two consecutive blobs), the size of the blobs, and the density of the blobs. The aim is to evaluate how a different fragmentation changes the structure of the post-shock region and the observables (namely
the distribution of emission measure vs. temperature and the line profiles). Note that the different parameter determine a different accretion rate (see Table~\ref{table}).}

\salvo{The effect of the blob frequency was investigated by comparing the reference case (run TR-D11-T300-R10.3) to two additional simulations: one with half a distance between two consecutive blobs (run TR-D11-T150-R10.3), and the other with twice a distance (run TR-D11-T600-R10.3). The density of the blobs is the same as in the reference simulation. As in the reference case, the blob impacts produce upflowing surges that hit the subsequent downfalling blobs. In the simulation TR-D11-T150-R10.3 the interaction of the surging upflow with the blob occurs closer to the stellar surface (h= 0.08$R_{\odot}$) than in the reference case due to the smaller distance between two consecutive blobs. The interaction produces a shock in the downfalling blob with temperature of $10^7K$. This shocked plasma flows downward with speeds of $\approx$ 300 km s$^{-1}$. In the case TR-D11-T600-R10.3 the surging upflows have more time to move upwards and cool down before hitting the subsequent blobs. In this case the impact occurs at distances from the stellar surface (h= 0.27$R_{\odot}$) larger than in the reference case. As a result, the post-shock region is more extended above the chromosphere and is characterized by a large range in plasma temperature (from $10^4$ to $10^7$ K) and by plasma with downfalling velocities of $\approx$ 250-300 km s$^{-1}$.}

\salvo{We explored the effect of the size of the blobs on the evolution and structure of the post-shock plasma by comparing the reference case with a simulation of blobs with a smaller size (run TR-D11-T300-R9.9). In the latter case the upflowing surges developed earlier and therefore occur more frequently than in the reference case due to the smaller size of the blobs. The temperature of the post-shock region is $\approx 8\times10^6$ K and the surges hit the subsequent blobs at $\approx 0.14 R_{\odot}$. The structure of the post-shock plasma is analogous to that of the reference case: the impacts of upflowing surges and downfalling blobs generate shocks with temperature of $\approx 10^7$K and plasma  which fall down with velocities of $\approx$ 150-300 km s$^{-1}$.}

\salvo{The effect of blob density on the structure of the post-shock region was investigated by comparing the reference case with a simulation assuming denser blobs ($\rho = 5\times 10^{11}$ cm$^{-3}$; run TR-D11.7-T300-R10.3). The density of the interblob region and the blob
frequency are the same as those of the reference case. As expected, denser blobs penetrate more deeply in the chromosphere due to their higher ram pressure and the stand-off height of the shocks transmitted in the blobs is smaller than that in the reference case
(see \citealt{2010A&A...522A..55S}). Again upflowing surges develop after the shocks reach the upper boundary of the blobs and hit the following blobs generating plasma with temperature up to $\approx 10^7$K. The structure of the post-shock region is analogous to that found in
the other simulations and is characterized by plasma with temperatures in the range between $10^4$ and $10^7$ K and downfalling velocities of $\approx$ 300-400 km s$^{-1}$.}

\salvo{Finally, just as a guide, we performed an additional simulation describing the impact of a single blob with density $\rho=10^{11}$cm$^{-3}$, downfalling velocity 500  km s$^{-1}$, and size $1.87\times10^{10}$ cm (run 1BL-R10.3). This simulation can be considered as an extreme case in which the time between two consecutive blobs is much larger than the duration of observation. In this case, after the impact of the blob with the chromosphere, the resulting surge expands upward without any interaction with downfalling blobs. The post-shock plasma cools down rapidly and is expected to produce blueshift of emission lines to speeds around $\approx$ 400 km s$^{-1}$.}

\subsection{Randomly fragmented stream}
\label{sec:fragm}

\salvo{The more general case of a randomly fragmented stream was investigated through simulations describing a column with uniform density of $10^9$ cm$^{-3}$ and a series of circular blobs with random spatial
distribution and random density in the range $5\times10^{10}$ and $5\times10^{11}$ cm$^{-3}$.

In these simulations, the blob impacts reproduce all the cases explored assuming a train of blobs
and, in addition, consider the interaction of multiple shocks with downfalling fragments. We performed two simulations assuming the same mass accretion rate but with different granularity of stream fragmentation: run Frag-N20 considers 20 blobs with radius ranging between $3.48\times10^8$ cm and $1.39\times10^9$ cm (coarse fragmentation), and run Frag-N55 considers 55 blobs with radius ranging between $3.48\times10^8$ cm and $6.96\times10^8$ cm (fine fragmentation).}

\salvo{The evolution is quite similar in the two cases. As for the train of blobs (see Sect.~\ref{sec:train}), the first blobs impact onto the chromosphere and produce upflowing surges of post-shock plasma that expand through the interblob medium (see Fig. \ref{sim_frag} and on-line movie). The expansion ends when the surges hit the following falling blobs. At variance with
the case of a train of blobs, the interactions of the surges with the falling blobs occur at different altitudes due to the initial random positions of the blobs in the stream. As a result, the
structure of the post-shock region is very complex and consists of several knots and filaments of shock-heated plasma with a broad range of velocities, densities, and temperatures (see Fig.
\ref{sim_frag}). The two runs Frag-N20 and Frag-N55 differ mainly for the average extension of the post-shock region. In fact, in run Frag-N20 the average distance among the blobs is larger than that in run Frag-N55. As a consequence, the surges have more time to expand upwards, so that their interactions with the downfalling blobs occur, on average, at higher altitudes. Because of this longer expansion, we also expect that a plasma component contributing to a blue shift of emission lines (namely that due to the upflowing surges) would be more important in run Frag-N20 than in run Frag-N55.}

Our case of a randomly fragmented stream  is similar to the HD model developed by \cite{2014ApJ...797L...5R}. The main differences are that the latter, neglect the magnetic field and consider blobs with not perfectly vertical trajectory with respect to the chromosphere. As a result, in the case explored by \cite{2014ApJ...797L...5R}, the plasma dynamics is not influenced by the magnetic field lines and the small deviation from the vertical trajectory determines an asymmetric evolution of the post-shock plasma.
The surges that bounce back are not confined by the magnetic field and are free to expand upward and laterally, causing a more efficient adiabatic cooling of the plasma and a lower upflowing velocity. Also the inclined trajectory of the blobs causes the surges to develop preferentially in the direction opposite to that of arrival of the blobs. For all of these reasons the post-shock region is very complex and characterized by a wide range of velocities of the shocked plasma.
 
\subsection{Distribution of emission measure vs temperature}
\label{sec:emt}

The distribution of emission measure versus temperature, EM(T), is a source of information of the plasma components \salvo{with different temperature} contributing to the emission\salvo{. In particular, we are interested on the components with temperatures $\log(T) \approx 5$~K, responsible of the C\,IV line emission, and with $\log(T) \approx 6.5$~K, contributing to the O\,VIII line.}

\begin{figure}
	\centering
		\includegraphics[scale=0.5]{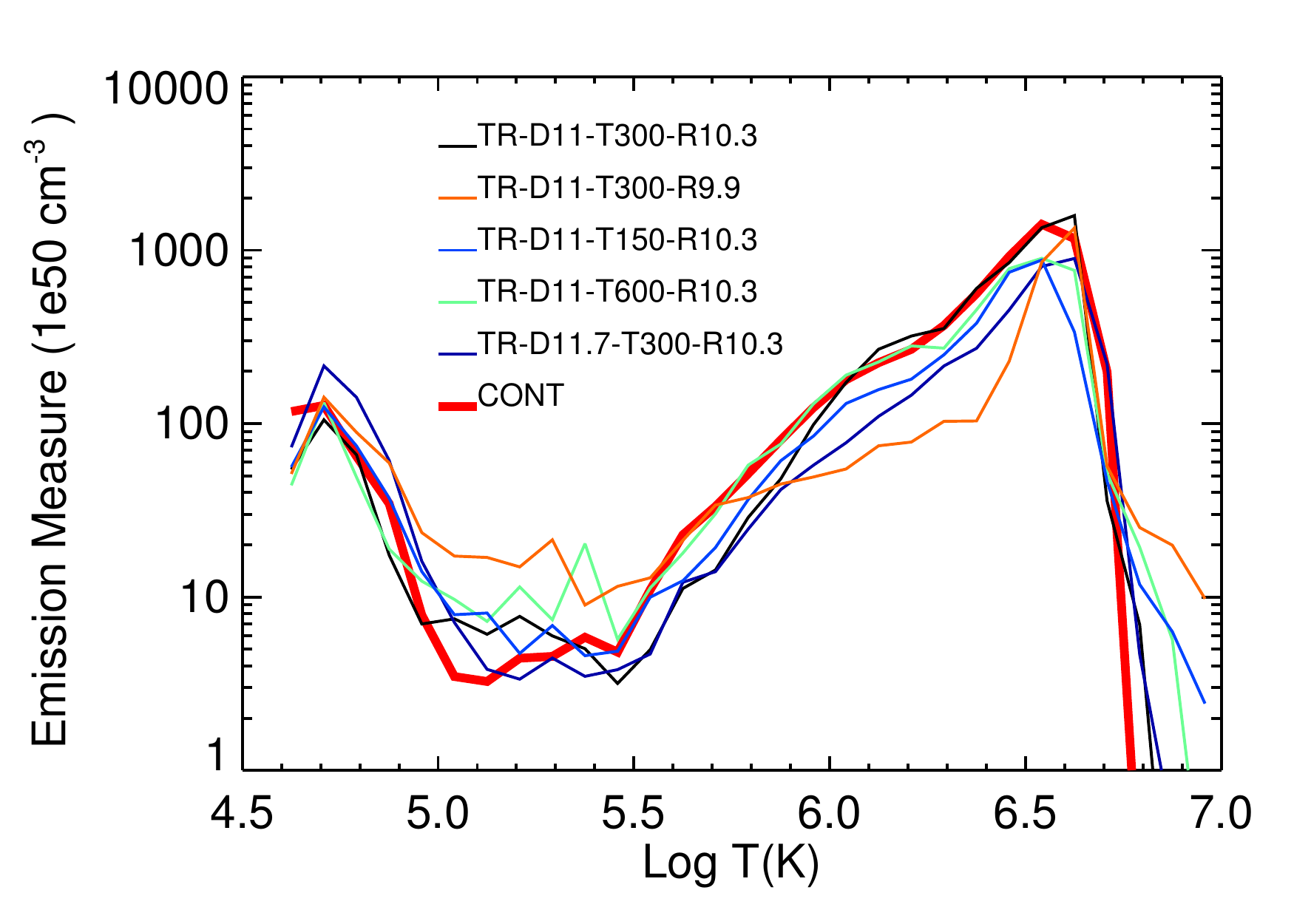}
	\centering
	    \includegraphics[scale=0.5]{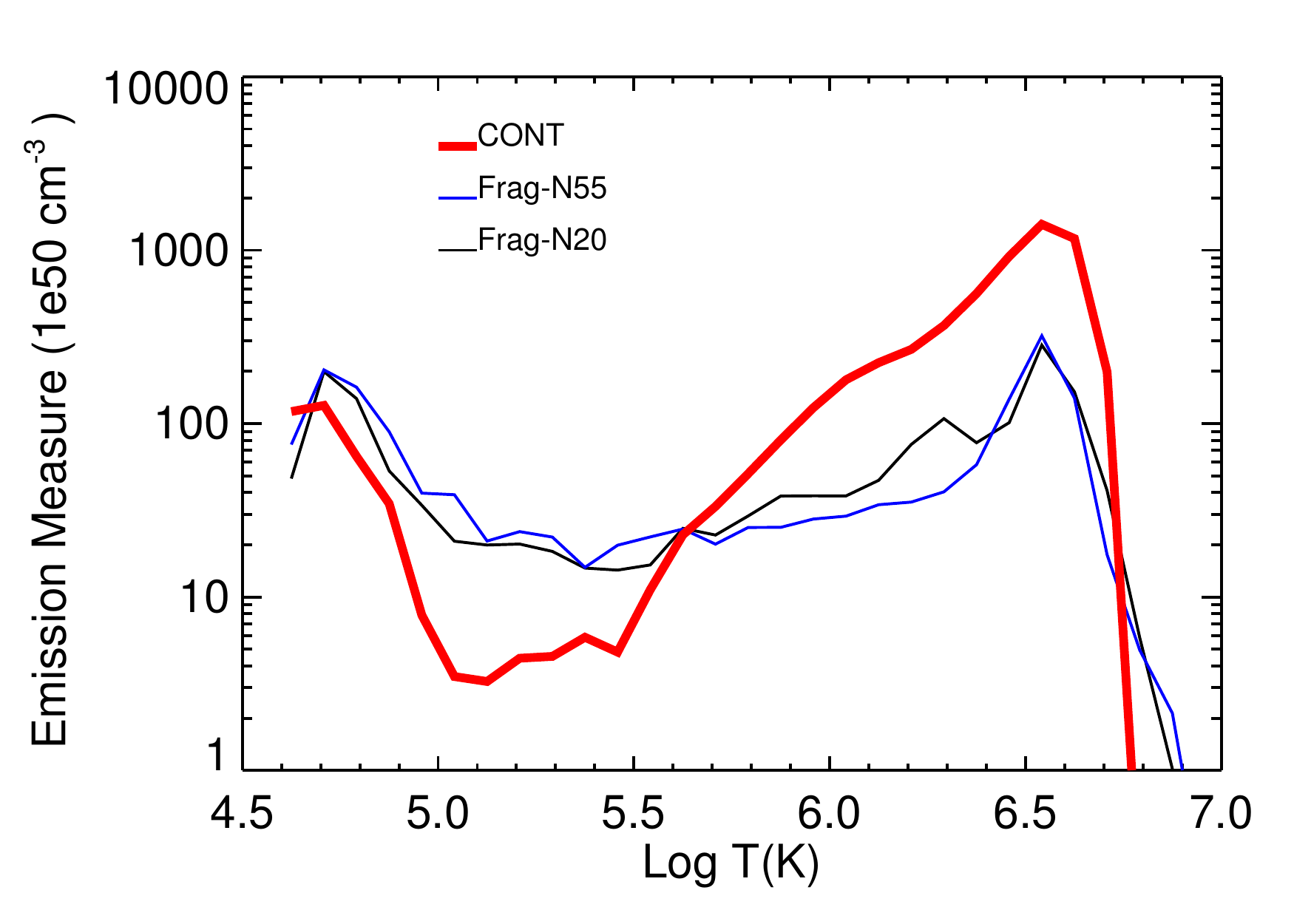}

\caption{\salvo{Distributions of emission measure vs temperature normalized to the accretion rate of TW Hya. The upper panel shows the distributions derived for a train of blobs and for a continuous stream; the lower panel shows the distributions derived for the general case of a fragmented stream.}}
\label{EM}
\end{figure}

\salvo{From the model results, we derived the EM$(T)$ distributions, following the methodology outlined in Sect.~\ref{sec:synth}. The distributions are obtained for all the frames of our simulations and then averaged over the time covered by the simulations. Then they are rescaled to a mass accretion rate of $10^{-9.17}$M$_\odot$ yr$^{-1}$ in agreement with  optical observations of TW~Hya \citep{2011A&A...526A.104C}.
The upper panel of Fig.~\ref{EM} shows the result for the trains of blobs. All these EM$(T)$ distributions have a shape similar to that of the continuous stream (red curve in the figure) and show two peaks, one at $\log T (K) \approx 4.7$ and the other at $\log T (K) \approx 6.5$. The former is due to the shocked plasma that cools down catastrophycally, the latter to the shock-heated plasma (see also \citealt{2010A&A...522A..55S}). The distributions also present a pronounced dip at $\log T (K) \approx 5-5.5$ (namely in the range of temperatures where the radiative losses are more efficient) in which the EM can be up to 2 orders of magnitude lower than that at $\log T (K) \approx 6.5$. By comparing the simulations of the train of blobs with that describing a continuous stream, we note that the former in general have larger emission measure in the range of temperatures between $\log T (K) \approx 5$ and $\log T (K) \approx 5.5$ and smaller emission measure at $\log T (K) \approx 6.5$ than the continuous case.}

\salvo{The lower panel of Fig.~\ref{EM} shows the EM$(T)$ distributions derived from the simulations describing the randomly fragmented stream (runs Frag-N20 and Frag-N55). Again the distributions present two peaks at $\log T (K) \approx 4.7$ and $\log T (K) \approx 6.5$. But the distributions are flatter in the temperature range between $\log T (K) \approx 5$ and $ \log T (K) \approx 6.5$, and the dip at $\log T (K) \approx 5.5$ is shallower to less	 than one order of magnitude from the peak at $\log T (K) \approx 6.5$. In fact, in these simulations, the fine stream fragmentation makes the impact frequency higher than in the case of a train of blobs. As a result, the post-shock region is quite complex and characterized by several structures of shock-heated plasma interacting each other and with a broad range of densities and temperatures (see Sect.~\ref{sec:fragm}). Even if the radiative losses are very efficient around $\log T (K) \approx 5$, several plasma structures which cool down are present at the same time in that range of temperatures thus contributing to increase the EM(T) values there.}

\subsection{UV and X-ray emission}

\begin{figure}[htbp]
		
		\centering
		\includegraphics[scale=0.5]{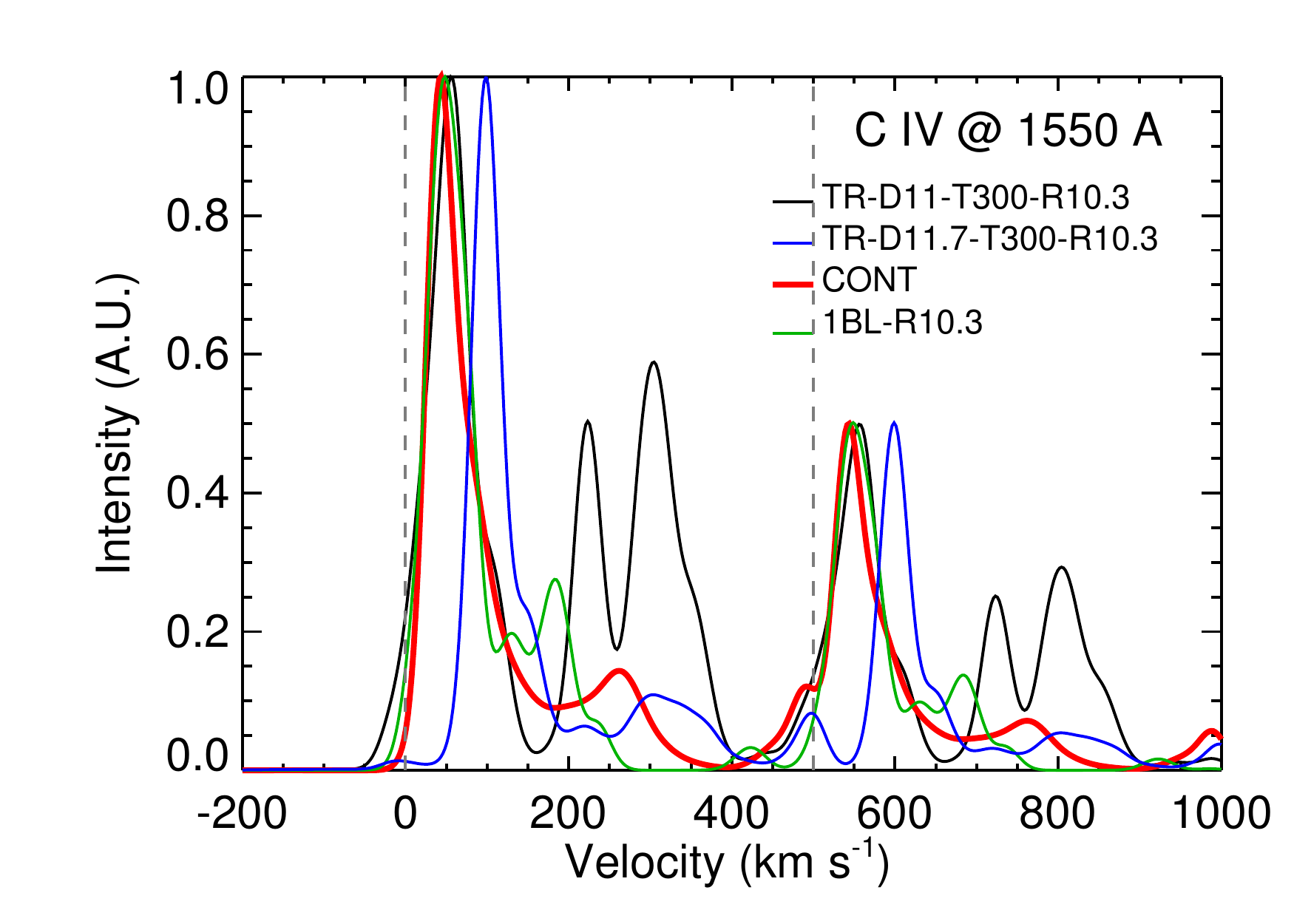}
		\includegraphics[scale=0.5]{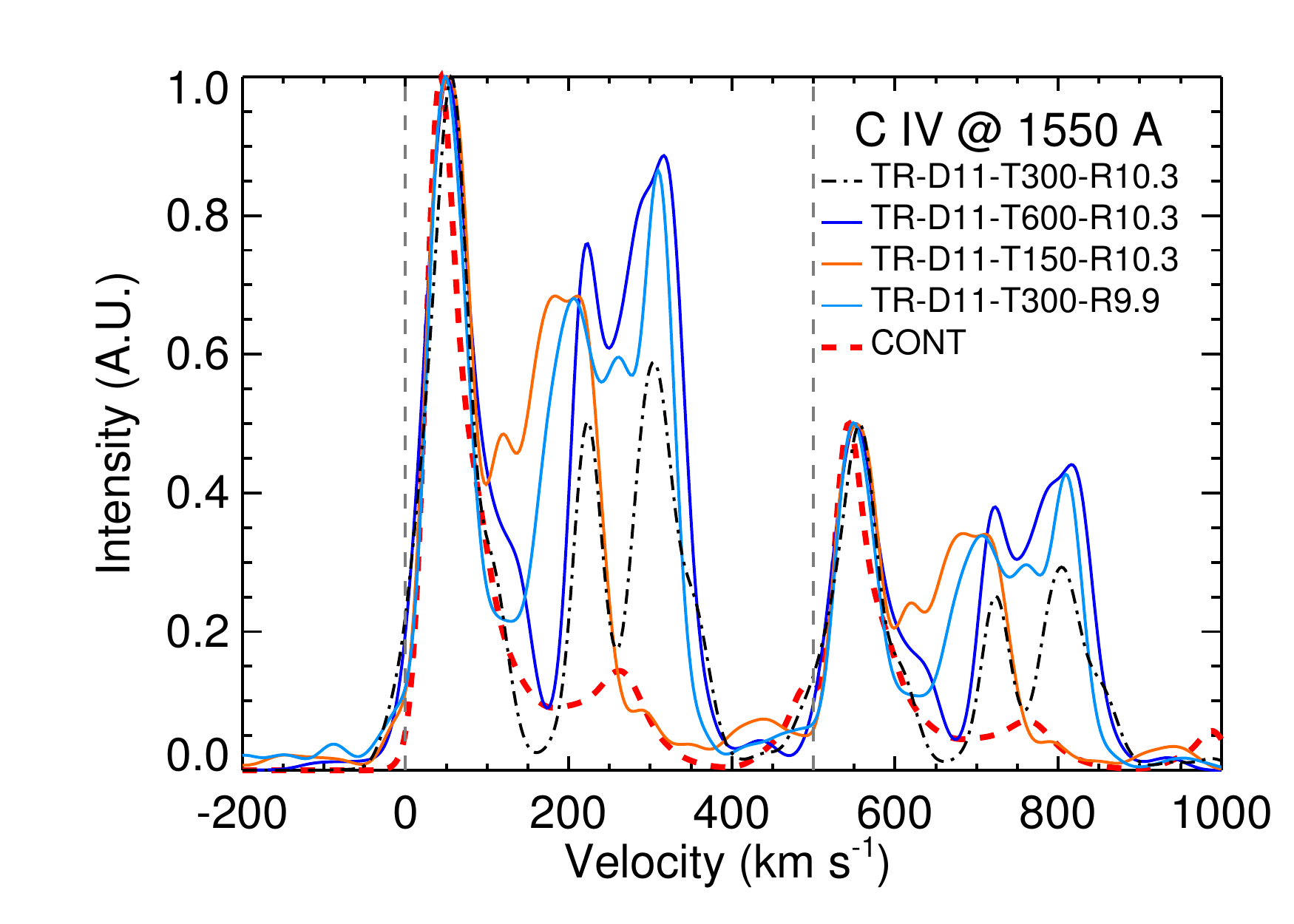}
	\caption{\salvo{Synthetic profiles} of C\,IV doublet; all
	the \salvo{profiles} are normalized to their maximum value.
	The \salvo{upper} panel \salvo{compares our reference
	simulation of a train of blobs (run TR-D11-T300-R10.3) with the case
	of a continuous stream (run CONT), the case of a single
	downfalling blob (run 1BL-R10.3), and with the case of a train of
	blobs with higher density (run TR-D11.7-T300-R10.3). The lower panel
	compares profiles derived from simulations of a train of
	blobs differing for the frequency and size of blobs (runs
	TR-D11-T300-R10.3, TR-D11-T600-R10.3, TR-D11-T150-R10.3, TR-D11-T300-R9.9).} The dotted grey
	lines are the rest positions of the lines of the
	doublet.\label{Spec}}
\end{figure}

\salvo{From the model results, we synthesized the emission in the UV and X-ray bands (see the method outlined in Sect.~\ref{sec:synth}), focusing our analysis on the doublets of C\,IV (1550 \AA) and
O\,VIII (18.97 \AA). In all the cases, we assume that the line of sight is aligned with the stream axis, thus maximizing the line shifts. Figure \ref{Spec} shows the profiles of the C\,IV doublet derived from models with trains of blobs. The line profile derived from the model of a continuous stream is also overplotted for comparison (red curve). Note that the line profiles are convolved with a Gaussian of $\sigma$ 17 km s$^{-1}$ (for C\,IV) to approximate the spectral resolution of HST \citep{2013ApJS..207....1A} observations. 
and 81 km s$^{-1}$ (for O\,VIII) to approximate the spectral resolution of Chandra/HETG.
}

\salvo{In the case of a continuous stream, the C\,IV line profile is characterized by a prominent slow component (hereafter SC) with a net redshift of $\approx 50$~km~s$^{-1}$ and by a second less intense and faster component (hereafter FC) with a redshift of $\approx 250$~km~s$^{-1}$. The former originates from the base of the accretion column close to the stellar chromosphere where the post-shock plasma cools down under the effect of thermal conduction and radiative losses and decelerates from $\approx 120$ to few km~s$^{-1}$.  The FC originates from catastrophic cooling of dense plasma accelerated to velocities around 200-250 km~s$^{-1}$ by the downfalling plasma above. We note also that the simulation of a single blob (run 1BL-R10.3) shows a profile analogous to that of a continuous stream. As for the latter, in fact, the contribution to the C\,IV line originates mainly from the cooling plasma at the base of the blob (responsible of the SC) and from the post-shock plasma subject to catastrophic cooling (responsible of the FC).}

\salvo{For a train of blobs, the line profiles are
much more complex than for a continuous stream due to
the more structured post-shock region. The lines
exhibit broadening and asymmetries reflecting the plasma structures
with different downfalling velocities present in the post-shock region. As for the case of a continuous stream, all
the lines are characterized by a SC. Again this component is due
to post-shock plasma which cools down and decelerates at the base
of the accretion column. The main differences with the case of a
continuous stream are in the FC. In the train of blobs, the FC
can be large with intensity even comparable with that of the SC. In general the FC is split
into multiple components with redshift ranging between
200 and 400 km~s$^{-1}$ (see Fig.~\ref{Spec}). {These components originate from the interaction of upflowing surges with downfalling blobs and from plasma subject to catastrophic cooling induced by radiative losses (see $\Lambda(T)$ in Eq.\ref{RL}) , as in the case of a continuous stream . }
Our simulations show that the relative intensity of the FC increases when the surges have more time to expand upward (for instance in runs TR-D11-T600-R10.3 and TR-D11-T300-R9.9), suggesting that the finer the fragmentation, the higher is the intensity of the FC. Also we note that the FC is broader for the train of blobs than for the continuous stream because of the larger range of velocities of plasma structures contributing
to C\,IV emission in the former case. Finally we note that both the
SC and FC are more redshifted for higher density of the blobs (see
upper panel in Fig.~\ref{Spec}) and the FC is less intense than in
the other cases.}

\begin{figure}

	\centering	
	\includegraphics[scale=0.5]{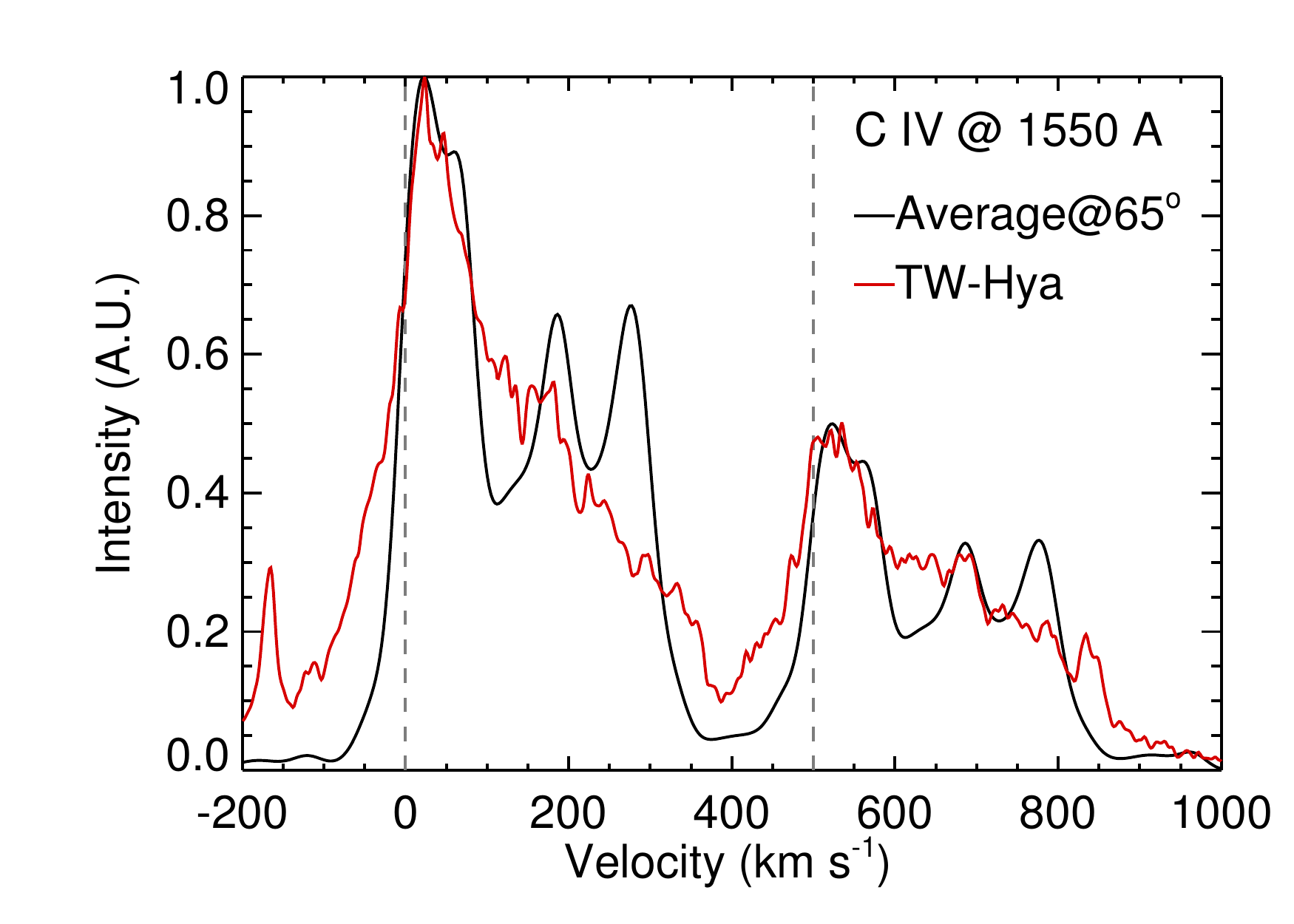}
	\caption{{Comparison between the synthetic profile
	of C\,IV doublet resulting from summing all the line profiles
	derived from the simulations of a train of blobs (black)
	and the observed spectrum of TW Hya analyzed by
	\cite{2013ApJS..207....1A} (red). We assumed an angle between
	the line of sigth and the axis of the accretion column of
	65 degrees to fit the velocity shift of the peak emission
	of the observed line. The profiles are normalized to their
	maximum value. The dotted grey lines represent the rest
	positions of the doublet.\label{media_civ}}}
\end{figure}

{From all the simulations of a train of blobs, we derived
the synthetic C\,IV line profile resulting from summing all the
line profiles normalized to the accretion rate of TW Hya.
This profile is expected either from a single accretion stream consisting of a train of blobs with different size and density or from few streams each with different train of blobs and accreting onto the CTTS at the same time.
In Fig. \ref{media_civ} we compare this synthetic C\,IV line profile 
with the profile observed from TW Hya and analyzed by
\cite{2013ApJS..207....1A} both normalized to their maximum. 
We also assumed an angle of 65 degrees between the line of sight and the axis of the accretion column to fit the velocity shift of the emission maximum of the observed profile.
This angle is in good
agreement with that suggested in the literature (e.g.
\citealt{2013ApJS..207....1A}). The synthetic profile can be described by
three main components: a narrow slow component with a redshift of
$\approx 50$~km~s$^{-1}$ and two faster components with a redshift
ranging between $200$ and $250$~km~s$^{-1}$. Note that the observed profile shows only one broad fast component that we interpret as the envelope of several fast components (see below). In general, we found that the synthetic profile
reproduces the asymmetries observed in the profile from TW Hya. The slow component in the synthetic profile should correspond
to the observed narrow component found by \cite{2013ApJS..207....1A} (see Fig.
\ref{media_civ}). The two faster components may explain the origin
of the observed broad component. This
is especially true if we consider that, in a more realistic scenario,
few accretion streams with different viewing angles (and therefore
different line shifts) might coexist, so that the two faster
components would be broader and possibly merge into a single broad
fast component.}

\begin{figure}

	\centering
	\includegraphics[scale=0.5]{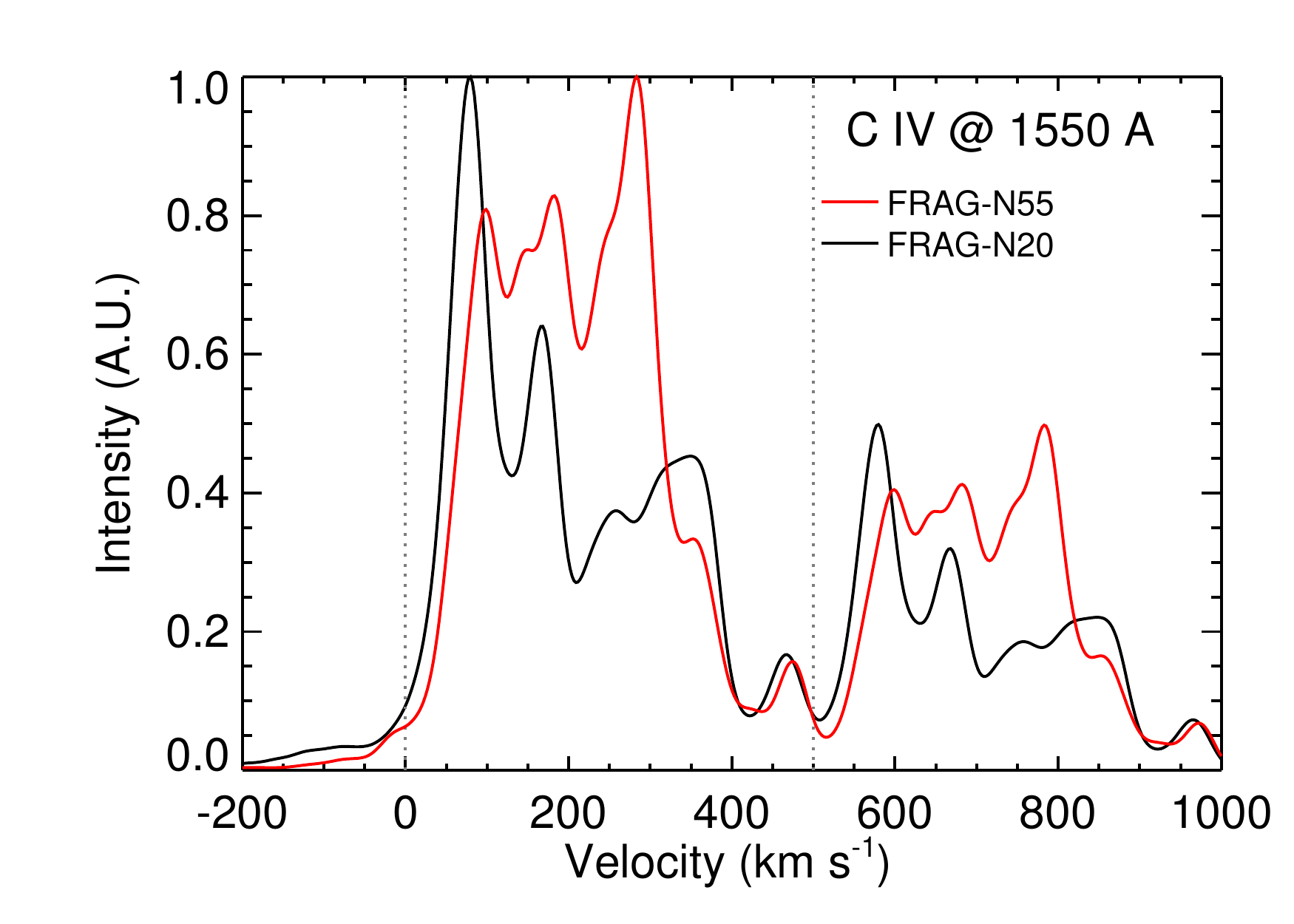}
	\caption{\salvo{As in Fig.~\ref{Spec} for the general case
	of a fragmented stream (runs Frag-N20 and Frag-N55).}}\label{spec_rand}
\end{figure} 

\salvo{Fig. \ref{spec_rand} shows the synthetic line profiles derived
from the randomly fragmented stream (runs Frag-N20 and
Frag-N55). As for the train of blobs,
 The line profiles are still characterized
by a narrow SC at $\approx 100$ km~s$^{-1}$ due to the cooling plasma
at the base of the accretion column. In the general case, however,
the profiles are more complex than for the train of blobs and show
several fast components with redshift ranging between 200 and 400
km~s$^{-1}$. Such a complexity is due to the structure of the
post-shock region which consists of several knots and filaments,
possibly subject to radiative cooling, with a broad range of
velocities. As expected in the light of the results obtained for a
train of blobs, the high level of stream fragmentation increases
the relative intensity of the faster components. An extreme case
is run Frag-N55 in which the most prominent component is at $\approx
300$~km~s$^{-1}$. In fact, as also discussed in Sect.~\ref{sec:emt},
the larger is the number of blobs with different density and size,
the more frequent the interactions of upflowing surges with downfalling
blobs, and the larger the number of structures of shock-heated plasma
with a broad range of densities, velocities and temperatures. Many of these
structures cool down under the effect of radiative losses, thus
contributing to the emission of the C\,IV doublet.}

\begin{figure}[!htbp]

\centering	
\includegraphics[scale=0.5]{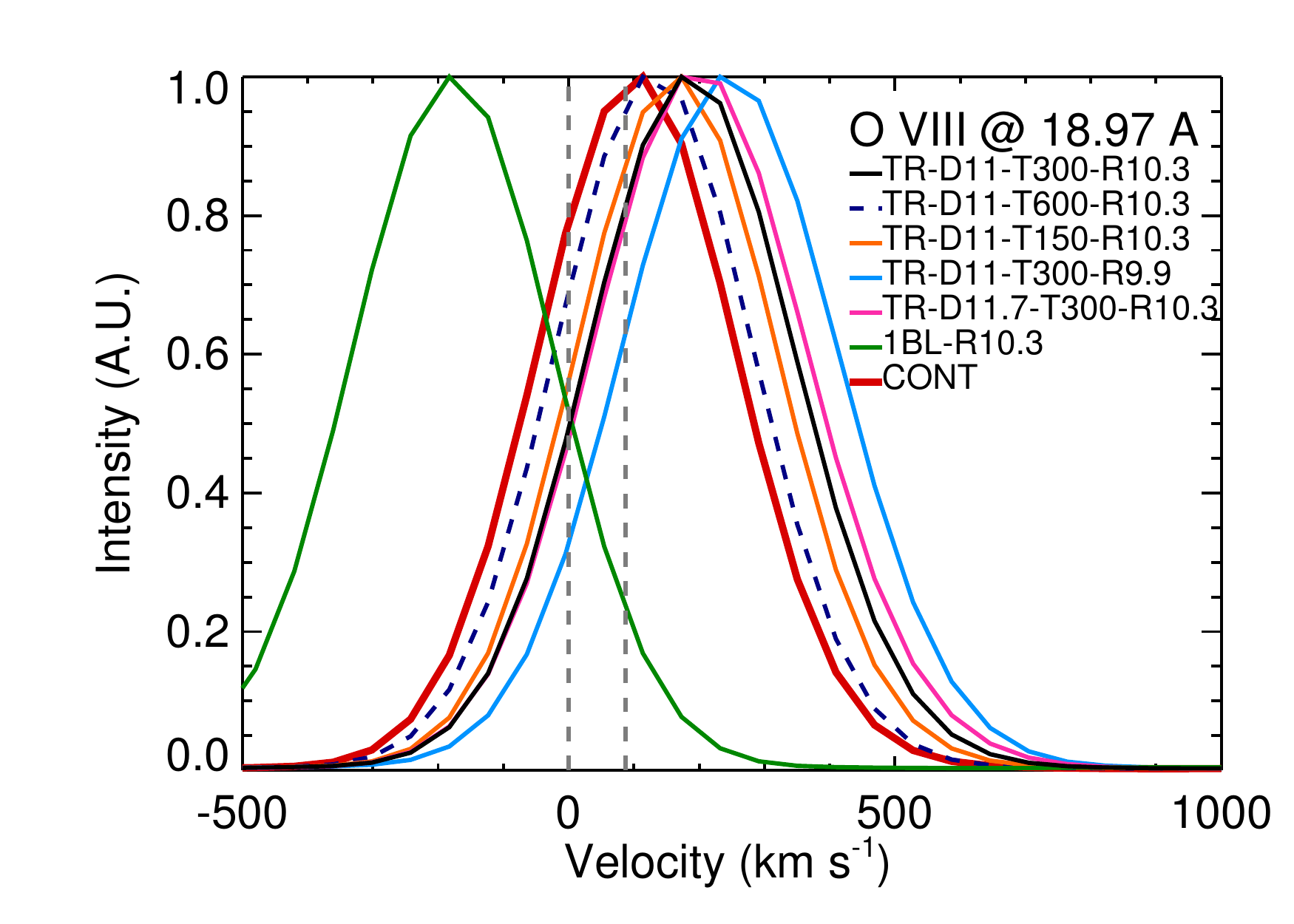}
\caption{\salvo{Synthetic profile of O\,VIII doublet normalized to
their maximum value for the simulations of a train of blobs, the
simulation of a single blob, and the case of a continuous stream.}
The dotted grey lines represent the rest positions of the doublet.
\label{OVIII}}
\end{figure}

\salvo{From the models, we synthesized also the O\,VIII doublet 
in the soft X-ray band. Figure \ref{OVIII} shows the line profiles
for the simulations of a train of blobs and for a continuous stream,
assuming again the line-of-sight aligned with the stream axis. For O\,VIII, the profiles are simpler than those of C\,IV mainly
because of the lower spectral resolution of available X-ray
instruments. However, despite the low resolution, some information
can be obtained from the analysis of O\,VIII lines. In fact, the
figure clearly shows that, in principle, a net redshift due to the accretion process is detectable\footnote{However, note that the shifts
presented here are the maximum expected for a downfall velocity of
500 km~s$^{-1}$, being the line of sight aligned with the stream axis.}, depending on line S/N, and viewing direction.
The case of a continuous stream (red curve in the figure) presents
the lines with the minimum redshift, with a velocity of $\approx 100$~km~s$^{-1}$,
compatible with the post-shock velocity. The simulations of a train
of blobs present higher redshifts because of the contribution of
hot plasma at the interaction region between the upflowing surges
and the downfalling blobs. The redshift increases roughly with the
level of stream fragmentation; the highest redshift is found for
run TR-D11-T300-R9.9, namely the case assuming smaller blobs. Finally,
a net blueshift ($\approx 200$~km~s$^{-1}$) is found for the case
of a single blob (run 1BL-R10.3), for which the
emission arises mainly from the post-shock plasma free to expand
upward. It should be noted, however, that the line intensity derived for
this case is more than one order of magnitude lower than those
derived in the other cases (Fig. \ref{OVIII} shows intensities normalized to the maximum).
}
\begin{figure}
	\centering
	\includegraphics[scale=0.5]{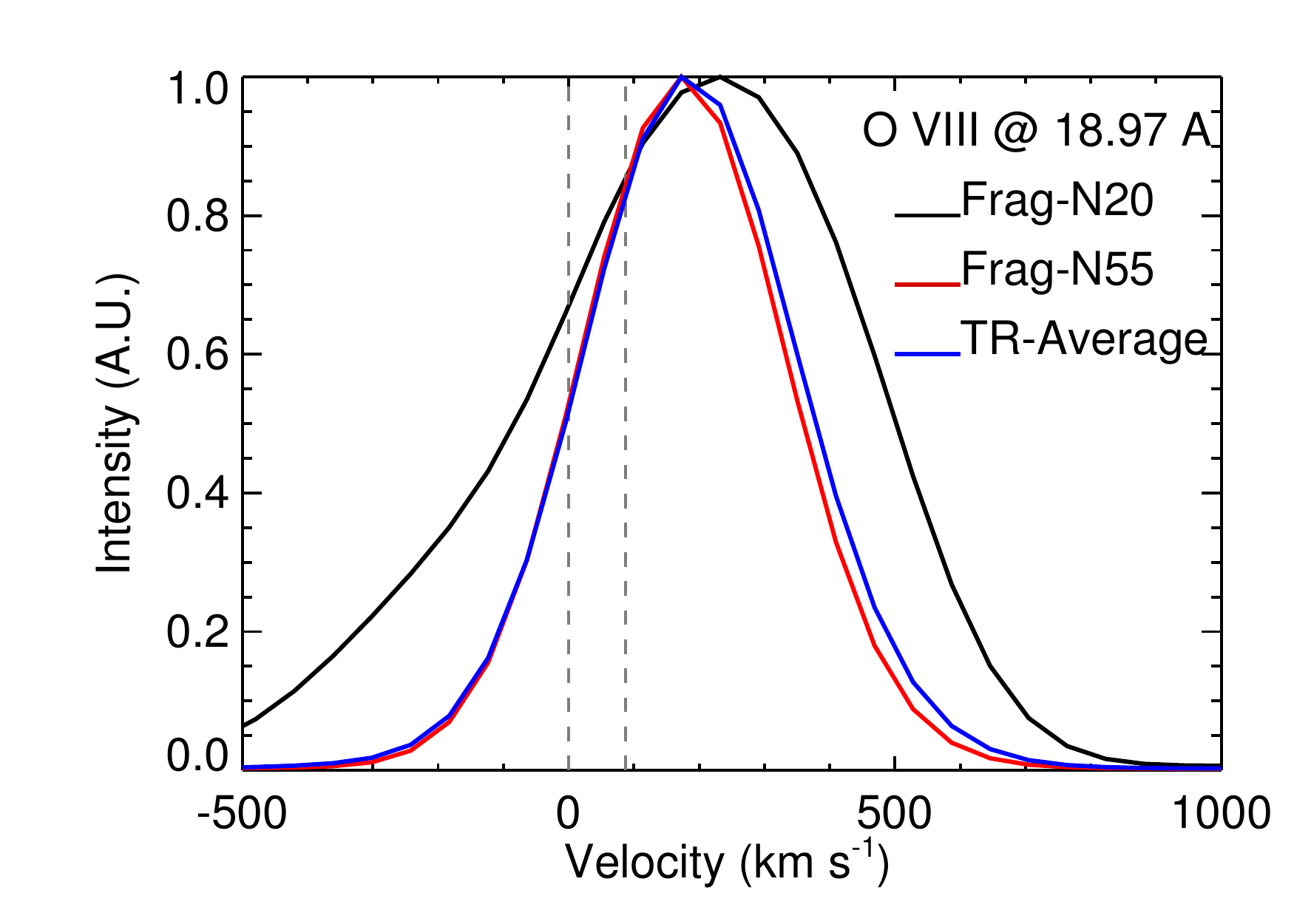}
	\caption{\salvo{As in Fig.~\ref{OVIII} for the randomly fragmented stream (runs Frag-N20 and
	Frag-N55). TR-Average is the average line profile obtained from the train of blobs simulations.}}\label{rand_oviii}
\end{figure}

\salvo{The O\,VIII doublets synthesized for the case of a randomly
fragmented stream (runs Frag-N20 and Frag-N55) are shown in Fig.
\ref{rand_oviii}. The results of run Frag-N55 are similar to those
derived for the train of blobs. The line appears symmetric and
redshifted to speed of $\approx 200$~km~s$^{-1}$. The profile almost
superimpose to that obtained summing all the spectra derived from
the simulations of a train of blobs (blue curve in
Fig. \ref{rand_oviii}). On the contrary, the line profile synthesized
from run Frag-N20 is asymmetric (with an intense and extended blue
wing) and broader than in run Frag-N55. We note that the blobs in
run Frag-N20 are sparser than in run Frag-N55. As a result, the surges
have more time to expand upward without impacts with downfalling
blobs, thus contributing more to the blue wing of
the O\,VIII line.}

\begin{figure}
	\centering
	\includegraphics[scale=0.5]{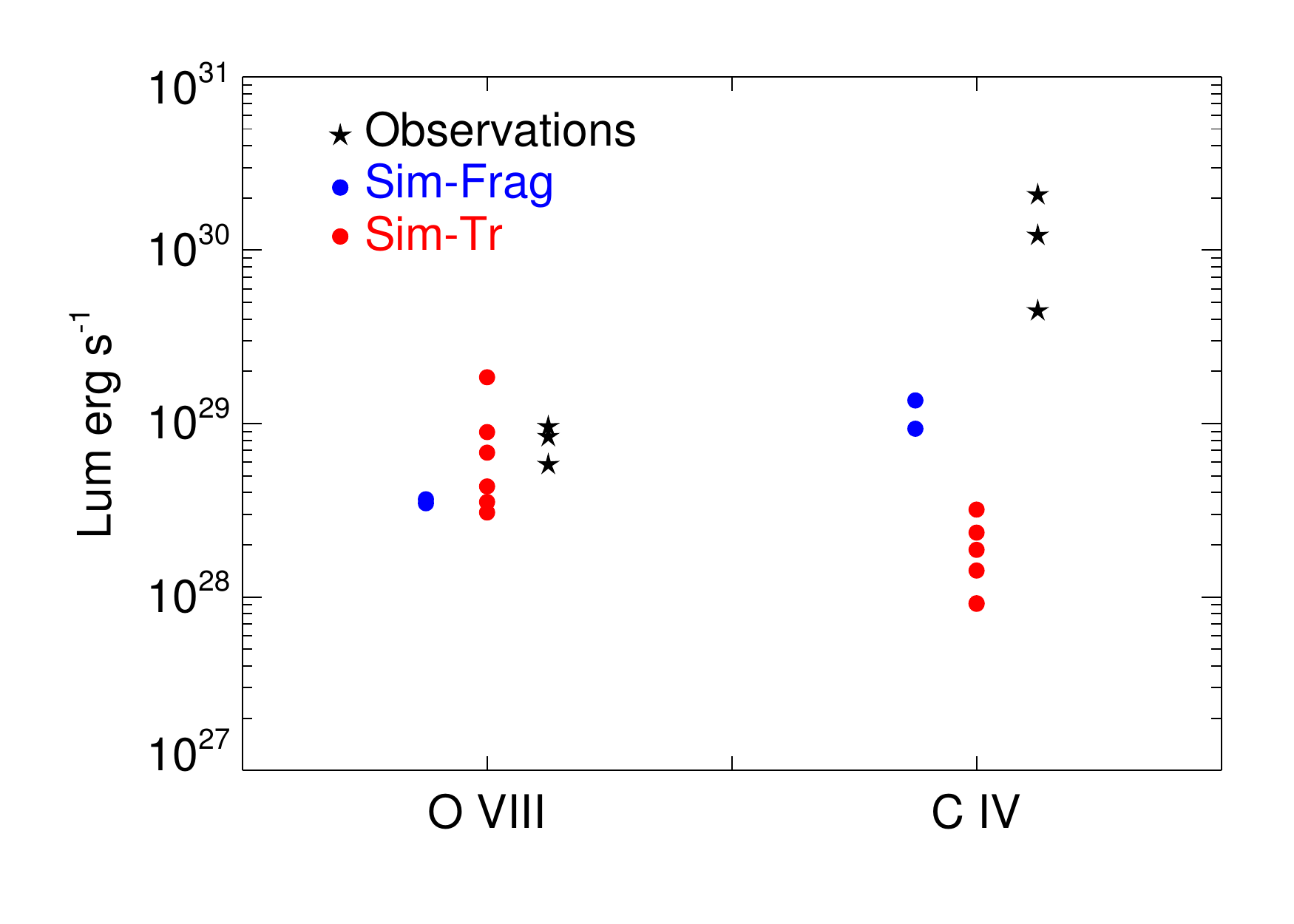}
	\caption{\salvo{Luminosity of C\,IV and O\,VIII lines
	synthesized from the models for the case of a train of blobs
	({red} dots; Sim-Tr) and for the case of randomly fragmented stream
	(blue dot; Sim-Frag). The luminosity derived from UV and X-ray
	observations of three well-studied CTTSs are superimposed
	({black} dots; MP~Mus, \citealt{2007A&A...465L...5A}; TW~Hya,
	\citealt{2010ApJ...710.1835B}; V4046~Sgr,
	\citealt{2012ApJ...752..100A}).}}\label{lum}
\end{figure}

\salvo{Finally, we derived the luminosity of C\,IV and O\,VIII
doublets from all our simulations and compared the synthetic values
with those derived from UV and X-ray observations of three well-studied
CTTSs, namely MP~Mus (\citealt{2007A&A...465L...5A}), TW~Hya
(\citealt{2010ApJ...710.1835B}), and V4046~Sgr
(\citealt{2012ApJ...752..100A}). The results of our simulations and
the comparison with observations are shown in Fig.~\ref{lum}. The
models are normalized in order to have a mass accretion rate of
$10^{(-9.17)}$ M$_\odot$ yr$^{-1}$ (namely comparable with those inferred from optical observations
of TW Hya; Curran et al. 2010). We found that, in general, our models
reproduce quite well the order of magnitude of the observed O\,VIII luminosity. This was
expected for the case of a continuous stream \citep{2010A&A...522A..55S}; we show that models of fragmented streams predict
luminosities of the O\,VIII doublet a bit lower than those for models
of continuous streams. On the other hand, we found that models of
a train of blobs predict C\,IV luminosities significantly lower
than those observed, although they are in better agreement with the
observations than models of a continuous stream (see Fig.~\ref{lum}).
Recently Costa et al. (2016, submitted) have proved that such a discrepancy
may be reconciled if the effect of the radiative heating of the
infalling material by the post-shock plasma is taken into account
in the models. Nevertheless, it is interesting to note that, in the
case of highly fragmented streams (runs Frag-N20 and Frag-N55), 
our model predict C\,IV luminosities significantly higher than those found by a train of blobs.
Inspecting Fig. \ref{lum} we note that normalizing the X-ray luminosity synthesized from runs Frag-N20 and Frag-N55 to observations also the UV luminosity synthesized by the same models is shifted at higher values, thus providing a better agreement.
(see Fig.~\ref{lum}).
 This is also in agreement with the results
of Sect.~\ref{sec:emt}: the EM$(T)$ distributions are characterized
by a lower dip at $\log(T)\approx 5$~K (namely around the temperature
of formation of the C\,IV doublet) if the stream is highly fragmented.
In fact, in these cases, a variety of plasma structures with different
density and temperature characterizes the post-shock region, most
of them cool down because of radiative losses and contribute to the
emission of the C\,IV doublet. We suggest that a high stream
fragmentation may contribute to increase the UV emission of accretion
impacts.}

\section{Summary and conclusions}
\label{sec4}

In this work we investigated the effects of \salvo{stream fragmentation}
on C\,IV and O\,VIII emission lines. We developed a model
that describes \salvo{a stream composed by a series of dense blobs}
impacting onto the surface of a CTTS. Our model \salvo{takes} into
account the stellar magnetic field, the gravity, the radiative
\salvo{losses} from optically thin plasma and the thermal conduction.
\salvo{The aim was to explore if and how the impact of a fragmented
stream onto the chromosphere of a CTTS reproduces profiles of C\,IV
doublet similar to those observed by \cite{2013ApJS..207....1A}.
Our main findings can be summarized as follow:}

\begin{itemize}
\item \salvo{The impact of a series of dense blobs onto the stellar
surface produces a more complex post-shock region 
than that in the case of a continuous stream. In particular the
blob impacts produce strong shocks propagating through the blobs and
then upflows after the blobs are fully shocked. The upflows in turn
hit and shock the still downfalling blobs, producing a large variety
of plasma structures (knots, filaments) differing in density,
temperature, and downfalling velocity. These structures are not present in the
case of a continuous stream.}

\item \salvo{If the stream is fragmented the C\,IV (1550 \AA) lines have a highly
asymmetric and broad profiles. The lines split into a narrow and intense component
redshifted to speed $\approx 50$~km~s$^{-1}$ and a multitude
of faster components redshifted to speeds in the range between 200
and 400~km~s$^{-1}$. The narrow component originates from the 
post-shock plasma at the base of the accretion column which cools
down under the effect of thermal conduction and radiative losses.
The fast components originate from thermal instabilities occurring
at high altitudes in the shocked stream and from the plasma structures
forming during the interaction of upflowing surges and downfalling
blobs. This is in agreement with \cite{2014ApJ...797L...5R}.}

\item \salvo{The intensity and velocity of the fast components
depend on the stream fragmentation: the finer is the fragmentation, the more intense are the fast components. Assuming a
more realistic scenario in which few accretion streams with different
viewing angles (and therefore different line shifts) are present,
the fast components easily merge together to form a single broad
component redshifted to speed $\approx 250$~km~s$^{-1}$. A similar result has been found by \cite{2014ApJ...797L...5R} by adopting an HD model to study blob falls on the solar surface. The narrow
component at $\approx 50$~km~s$^{-1}$ and the broad component at
$\approx 250$~km~s$^{-1}$ are analogous to those found by
\cite{2013ApJS..207....1A} for most of the CTTSs of their sample. Thus we interpret
the shape of observed C\,IV lines as evidence of density structured or
fragmented accretion streams.}

\item \salvo{The O\,VIII (18.97 \AA) lines have a symmetric observable profile
redshifted to speed ranging between 100 and 200~km~s$^{-1}$. The
redshift increases roughly with the level of stream fragmentation:
the shift is the smallest in the case of a continuous flow and the
largest for a train of small blobs. In any case our model predicts
that accretion impacts would produce detectable shifts in the O\,VIII
lines.}

\item \salvo{As in the case of continuous accretion streams, models
of fragmented streams reproduce quite well the luminosity of O\,VIII
lines measured in CTTSs \citep{2007A&A...465L...5A,2010ApJ...710.1835B,2012ApJ...752..100A} and, in general, underestimate even by
orders of magnitude the luminosity of C\,IV lines. On the other
hand, we found that assuming an high level of stream fragmentation is in better agreement with observations. In these
models, in fact, many interactions between upflowing surges and
downfalling blobs are present which produce a multitude of plasma
structures with different density and temperature. Many of them cool
down because of radiative losses thus contributing to emission in
C\,IV lines. We conclude that the stream fragmentation enhances the
emission in the UV band.}
\end{itemize}

In conclusion, our models reproduce profiles of C IV and O VIII lines remarkably similar to those observed (\citealt{2013ApJS..207....1A}, Argiroffi in preparation) and suggest that the UV emission originates mainly from plasma structures developed as a result of the impact of a density structured or fragmented accretion stream. On the other hand our models predict in general UV luminosities lower than observed.
We note that oru models assume that the plasma is optically thin in the whole domain. However, optically thick plasma, as that of the chromosphere and of the unshocked stream, is present around the impact region.
 This plasma, on one hand, absorb part of the X-ray emission arising from the post-shock plasma \citep{2010A&A...522A..55S,2014ApJ...795L..34B} and, on the other hand, can be heated up to $ \log T (K) \approx 5$ by irradiation by the post-shock plasma, thus contributing to UV emission (Costa et al. 2016, submitted). In this work we did not take into account the absorption by optically thick material and the effect of radiative heating of the unshocked stream by post-shock plasma.
In a future work we plan to include these effects on the model to investigate more deeply the origin of UV emission arising from impact regions of fragmented accretion streams.
\begin{acknowledgements}
We are grateful to David Ardila fo providing the observed spectra of TW Hya.
We ackowledge support from INAF through the Progetto Premiale: "A Way to Other Worlds" of the Italian Ministry of Education, University, and Research.
PLUTO is developed at the Turin Astronomical Observatory in collaboration with the
Department of Physics of Turin University. 
We acknowledge also the CINECA Award HP10CWX941 and the HPC facility (SCAN) of the INAF – Osservatorio Astronomico di Palermo, for the availability of high performance computing resources and support. 
CHIANTI is a collaborative
project involving the NRL (USA), the Universities of Florence (Italy) and
Cambridge (UK), and George Mason University (USA).

\end{acknowledgements}
\bibliography{bib}
\bibliographystyle{natbib}

\end{document}